\documentclass{PoS}
\usepackage{amsmath}
\pdfoutput=1

\title{Determination of the running coupling in pure SU(4) Yang-Mills theory}

\ShortTitle{Running coupling in pure SU(4) Yang-Mills theory}

\author{Biagio Lucini\\
        Physics Department, Swansea University, Singleton Park, Swansea SA2 8PP, UK\\
        E-mail: \email{b.lucini@swansea.ac.uk}}

\author{\speaker{Gregory Moraitis}\\
        Physics Department, Swansea University, Singleton Park, Swansea SA2 8PP, UK\\
        E-mail: \email{pygm@swansea.ac.uk}}

\abstract{The Schr\"odinger functional is used to define a renormalised coupling for pure SU(4) Yang-Mills theory, with Wilson action and suitably selected boundary conditions on the link field. The coupling, which runs with the size of the lattice, is then determined by a finite-size scaling technique through a large range of momenta, thereby allowing a connection to be made between the high energy regime and the low energy non-perturbative regime, where contact is made with the physical scale of the theory. Using data from previous SU(2) and SU(3) simulations obtained with the same technique, the running of the 't Hooft coupling defined through the Schr\"odinger functional is studied, and we check whether the large-$N$ expectation that $\bar g^2N$ is a universal function of the energy scale $E$ holds down to energies of the order of the string tension. Finally, we determine $\Lambda_{SF}/\sqrt\sigma$ as a function of $N$ at leading order in $1/N^2$.}

\FullConference{The XXV International Symposium on Lattice Field Theory\\
                 July 30 - August 4 2007\\
                 Regensburg, Germany}

\begin{document}
\section{Introduction}
\label{sec:intro}
In perturbative calculations in QCD, dimensional transmutation is accounted for by a scheme-dependent quantity with mass dimension one called the $\Lambda$-parameter. Since QCD is a one-scale theory, $\Lambda$ must be related to the non-perturbative scale of the theory, which can then be used to express the perturbative running of the coupling. In this respect, the lattice has proved to be a valuable tool, and the running coupling has already been computed for pure SU(2) and SU(3) Yang-Mills theories~\cite{Luscher:1992zx,Luscher:1993gh} using a scheme referred to as the Schr\"odinger functional~\cite{Luscher:1992an}. Here, we apply this method to the SU(4) theory.

In the Schr\"odinger functional scheme, the renormalised coupling runs with the physical size of the system $L$. To cover a wide range of energies, an iterative step-scaling method is used~\cite{Luscher:1991wu}. A step function $\sigma(s,\bar{g}^2(L))=\bar{g}^2(sL)$ is introduced and determined pointwise for systems of size $L$ and $sL$ in every direction. To step up the scale, the bare coupling is tuned from $g_0$ to $g'_0$, such that $\bar g^2(L)|_{g'_0}=\bar g^2(sL)|_{g_0}$. This renormalisation group transformation can be performed iteratively and effectively allows us to accommodate larger system sizes on fewer lattice sites. Of course, once on a lattice, the system is affected by the finite lattice spacing $a$ so, in practice, each physical box size is simulated for different lattice spacings and an extrapolation to the continuum is performed. With this method, we successfully cover a range from 7GeV down to 0.44GeV.

At the lower end of this range, contact can be made with the non-perturbative observables of the theory. In particular, by using the data in~\cite{Lucini:2004my} and the interpolating formulae in~\cite{Lucini:2005vg}, we are able to express the coupling in physical units. Once the scale is fixed, we can obtain a value for $\Lambda_{SF}$ in the Schr\"odinger functional scheme, and compare the lattice data with the perturbative evolution of the beta function to two-loops.
\section{Method}
\label{sec:method}
\subsection{Preliminaries and notation}
\label{sec:notation}
The basic ideas and method are taken over directly from \cite{Luscher:1992an}, so only the basic notation is outlined below.

The Schr\"odinger functional $\mathcal{Z}$ is defined on the lattice with the standard Wilson action,
\begin{equation}\mathcal{Z}[C,C']=\int D[U]e^{-S[U]},\end{equation}
\begin{equation}S[U]=\frac{1}{g_0^2}\sum_{p}\operatorname{Tr}(1-U(p)),\end{equation}
where $U(p)$ denotes the plaquette, and the sum must be explicitly taken over both orientations.

The boundary links are required to satisfy inhomogeneous Dirichlet boundary conditions for $k=1,2,3$,
\begin{equation}
\label{eq:links}
W(\mathbf x,k)|_{x^0=0}=\exp(aC_k(\mathbf{x})),\qquad W(\mathbf x,k)|_{x^0=L}=\exp(aC'_k(\mathbf{x})),
\end{equation}
where $C_k$ and $C'_k$ are spatial boundary fields which need to be chosen (cf. \S\ref{sec:fund}).

The fixed boundary conditions induce a colour background field $B$ into the system so that the effective action can be written as an asymptotic series
\begin{equation}
\Gamma[B]=g_0^{-2}\Gamma_0[B]+\Gamma_1[B]+g_0^2\Gamma_2[B]+\cdots.
\end{equation}
By introducing a dependence of the boundary links (and thus of the background field) on a real dimensionless parameter $\eta$, we can then define a renormalised coupling as
\begin{equation}
\label{eq:coupling}
\bar{g}^2=\frac{\Gamma'_0[B]}{\Gamma'[B]},\qquad\Gamma'[B]=\frac{\partial}{\partial\eta}\Gamma[B],
\end{equation}
for a particular choice of $\eta$. On the lattice, this quantity can be measured via Monte Carlo simulation, by calculating the expectation value of the observable
\begin{equation*}
\frac{\partial S}{\partial\eta}=-\frac{ia}{g_0^2L}\sum_{\mathbf x}\sum_{l=1}^3\left[(E_l(\mathbf x)+E'_l(\mathbf x))+(E_l(\mathbf x)+E'_l(\mathbf x))^\dagger\right],
\end{equation*}
\begin{equation}
\label{eq:observable}
E_l(\mathbf x)=\operatorname{Tr}\left[cW(\mathbf x,l)U(x+a\hat{l},0)U(x+a\hat0,l)^\dagger U(x,0)^\dagger\right]_{x^0=0},
\end{equation}
where $c$ is a matrix appearing when differentiating the boundary field dependence of the Wilson action (cf. \S\ref{sec:fund}), and a similar expression holds for $E'(\mathbf x)$.
\subsection{The fundamental domain in the SU(4) theory}
\label{sec:fund}
It is desirable to choose boundary fields which minimise the effect of the finite lattice spacing. It was shown in~\cite{Luscher:1992an} that, for $N$ colours, the optimal choice are constant Abelian fields
\begin{equation}
C_k=\frac{i}{L}
 \left( \begin{array}{cccc}
                       \phi_{k1}  & 0       & \cdots & 0 \\
                           0      & \phi_{k2}  & \cdots & 0  \\
                           \vdots & \vdots  & \ddots & \vdots \\
                           0      & 0       & \cdots & \phi_{kN}
                        \end{array} \right), \qquad
C'_k=\frac{i}{L}
 \left( \begin{array}{cccc}
                       \phi'_{k1}  & 0       & \cdots & 0 \\
                           0      & \phi'_{k2}  & \cdots & 0  \\
                           \vdots & \vdots  & \ddots & \vdots \\
                           0      & 0       & \cdots & \phi'_{kN}
                        \end{array} \right).
\end{equation}
Stability considerations of the background field constrain the angles
\begin{equation}
\label{eq:conditions}
\sum_{i=1}^{N}\phi_i=0,\qquad\phi_1<\phi_2<...<\phi_N,\qquad|\phi_i-\phi_j|<2\pi
\end{equation}
and similarly for $\phi'$ (from here on we drop the suffix $k$ on the angles and use the same choice for $k=1,2,3$). A set of angles satisfying these constraints is said to be in the fundamental domain.

\begin{figure}
\begin{center}
\includegraphics[scale=1]{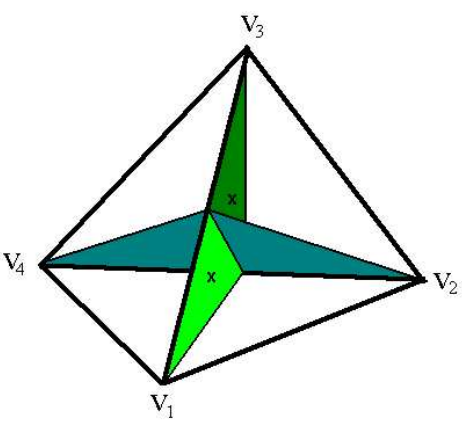}
\end{center}
\caption[The fundamental domain and its two planes of symmetry.]{The fundamental domain and its two planes of symmetry. The points marked $\mathbf x$ correspond to the angles \eqref{eq:angles}.}
\label{fig:domain}
\end{figure}

Specialising to SU(4), the fundamental domain can be described symmetrically by defining a one-to-one map between the set of angles $(\phi_1,\phi_2,\phi_3,\phi_4)$ and a point $\mathbf V$ in a certain bounded three-dimensional region,
\begin{equation}\mathbf{V}=\tfrac{3}{4}(\phi_1\cdot\mathbf e_1+\phi_2\cdot\mathbf e_2+\phi_3\cdot\mathbf e_3+\phi_4\cdot\mathbf e_4),\qquad\phi_i=\mathbf{V}\cdot\mathbf e_i,\end{equation}
where $\mathbf e_i$ are the weights of the Lie algebra of SU(4) in the fundamental representation, normalised as $\mathbf e_i\cdot\mathbf e_j=\tfrac{1}{3}(4\delta_{ij}-1)$. With this normalisation, the vertices $\mathbf v_i$ of the fundamental domain are
\begin{equation}\mathbf v_i=-\frac{3\pi}{2}\sum_{j=1}^{i}\mathbf e_j,\qquad i=1,2,3,4 \ ,\end{equation}
describing a skewed tetrahedron (figure \ref{fig:domain}).
Having identified the fundamental domain, one must then select the set of angles for each boundary, $C_k$ and $C'_k$, each set corresponding to a point inside the tetrahedron. These points will be members of a one-parameter family of angles, parameterised by $\eta$, through which the renormalised coupling is defined by \eqref{eq:coupling}. In principle this choice is arbitrary and of no conceptual significance, however the signal-to-noise ratio of the Monte Carlo evaluation is highest when (i) the points are well away from the domain's edge, (ii) they are as far as possible from each other, and (iii) the two boundaries are on an equal footing. Geometrically, this corresponds to choosing two points related by a symmetry of the fundamental domain.

Keeping the above points in mind, we choose the two points to be related by the symmetry reflecting about the plane through $\mathbf v_2$ and $\mathbf v_4$, and make the particular choice
\begin{align}
\label{eq:angles}
&\phi_1=-\tfrac{1}{2}\eta-\tfrac{1}{4}\pi\sqrt{2}     &\phi'_1&=\tfrac{1}{2}\eta-\tfrac{1}{4}(2+\sqrt{2})\pi \nonumber\\
&\phi_2=-\tfrac{1}{2}\eta-\tfrac{1}{4}(2-\sqrt{2})\pi &\phi'_2&=\tfrac{1}{2}\eta-\tfrac{1}{4}(4-\sqrt{2})\pi \nonumber\\
&\phi_3=\tfrac{1}{2}\eta+\tfrac{1}{4}(2-\sqrt{2})\pi  &\phi'_3&=-\tfrac{1}{2}\eta+\tfrac{1}{4}(4-\sqrt{2})\pi\nonumber\\
&\phi_4=\tfrac{1}{2}\eta+\tfrac{1}{4}\pi\sqrt{2}      &\phi'_4&=-\tfrac{1}{2}\eta+\tfrac{1}{4}(2+\sqrt{2})\pi.
\end{align}
With this choice, we set $\eta=0$ to obtain
\begin{equation}\Gamma'_0[B]=-\frac{24L^2}{a^2}\sin\left(\frac{a^2\pi}{2L^2}\right).\end{equation}
The matrix $c$ defined in \eqref{eq:observable} is then $c=\operatorname{diag}(-\tfrac{1}{2},-\tfrac{1}{2},\tfrac{1}{2},\tfrac{1}{2})$.
\section{Results}
\label{sec:results}
The results of our simulations are summarised in table \ref{tab:couplings}. The data is grouped into four blocks, each corresponding to a fixed value of $\bar g^2(L)$. We use a step of $s=2$ throughout, and calculate the renormalised coupling $\bar g^2(2L)$ for different lattice spacings, extrapolating to the continuum. The errors quoted are statistical only, and at this stage all results should be regarded as preliminary.
\begin{table}[tbp]
\begin{center}
\begin{tabular}{|lllll|l|}
	\hline
$\beta$ &  $L/a$    &   $a/L$  & $\bar{g}^2(L)$  & $\bar{g}^2(2L)$ & Extrapolation  \\
	\hline
15.126	& 6	  & 0.167 &	1.0222(6) &	1.2247(10)  &  \\
15.626	& 8	  & 0.125 & 1.0223(7)	& 1.2162(18)  & 1.1891(37)\\
16.000  & 10	& 0.100 &	1.0223(5)	&	1.2104(21)  & \\
 \hline
14.137	& 6  & 0.167 & 1.1893(3) & 1.4832(13) &\\
14.634	& 8	 & 0.125 & 1.1894(4) & 1.4700(19) & 1.4332(47)\\
15.007	& 10 & 0.100 & 1.1890(8) & 1.4634(25) &\\
 \hline
13.142 & 6   & 0.167 & 1.4328(3) & 1.9042(14) & \\
13.631 & 8   & 0.125 & 1.4332(4) & 1.8797(24) & 1.8099(62) \\
14.000 & 10  & 0.100 & 1.4330(6) & 1.8667(35) & \\
 \hline
12.190	& 6	  & 0.167 &	1.8098(7) &	2.7198(38) & 2.4601(300) \\
12.668	& 8	  & 0.125 & 1.8102(6)	& 2.6548(49) & \\
 \hline
\end{tabular}
\end{center}
\caption{Pairs of running couplings for fixed values of $\beta=8/g_0^2$.}
\label{tab:couplings}
\end{table}

\begin{table}[tbp]
\begin{center}
\begin{tabular}{|ll|}
	\hline
$L/a$ & $\beta$ \\
 \hline
6 & 11.329 \\
7 & 11.571 \\
 \hline
\end{tabular}
\end{center}
\caption{Bare coupling vs lattice sites at fixed $\bar g^2(L)=2.46$.}
\label{tab:bare}
\end{table}

To convert this dimensionless scale ($L/a$) to a more physical one, we need to set the lattice spacing at a fixed value of $\beta$. The quantity $a\sqrt\sigma$ is readily available in the range $10.55<\beta<11.40$~\cite{Lucini:2004my}, so we can deduce the physical lattice spacing at the highest couplings probed, ie. at $\bar g^2(L)=2.46$ (table \ref{tab:bare}). Using the interpolating formula in~\cite{Lucini:2005vg}, we obtain $a\sqrt\sigma=0.16069$ at $\beta=11.329$ and $a\sqrt\sigma=0.13523$ at $\beta=11.571$. These two values are used individually to compute the scale of the theory and the results match to better than 2\%. Using $\sqrt\sigma=420\operatorname{MeV}$, the maximum (minimum) box size is 0.45fm (0.028fm) corresponding to an energy range from 0.44GeV to 7GeV. From asymptotic freedom, we expect that perturbation theory holds accurately at the upper end of the scale, so we impose that $\bar g^2(7\operatorname{GeV})=1.0222$ and use the two-loop solution to the RG equation for the coupling $\bar g^2$ at an energy scale $E$
\begin{equation}
L=E^{-1}=\frac{1}{\Lambda_{SF}}\left(\frac{\beta_1}{\beta_0^2}+\frac{1}{\beta_0\bar g^2(E)}\right)^\frac{\beta_1}{2\beta_0^2}e^{-\frac{1}{2\beta_0\bar g^2(E)}}
\end{equation}
to obtain a value for the $\Lambda$-parameter for $N=4$ colours,
\begin{equation}\Lambda_{SF}(N=4)=101\pm1\operatorname{MeV}.\end{equation}
The error quoted only reflects the scale uncertainty which comes from fixing the lattice spacing at the two values of $\beta$, though this is expected to be the largest source of error.

\begin{figure}
\begin{center}
\includegraphics[scale=0.45]{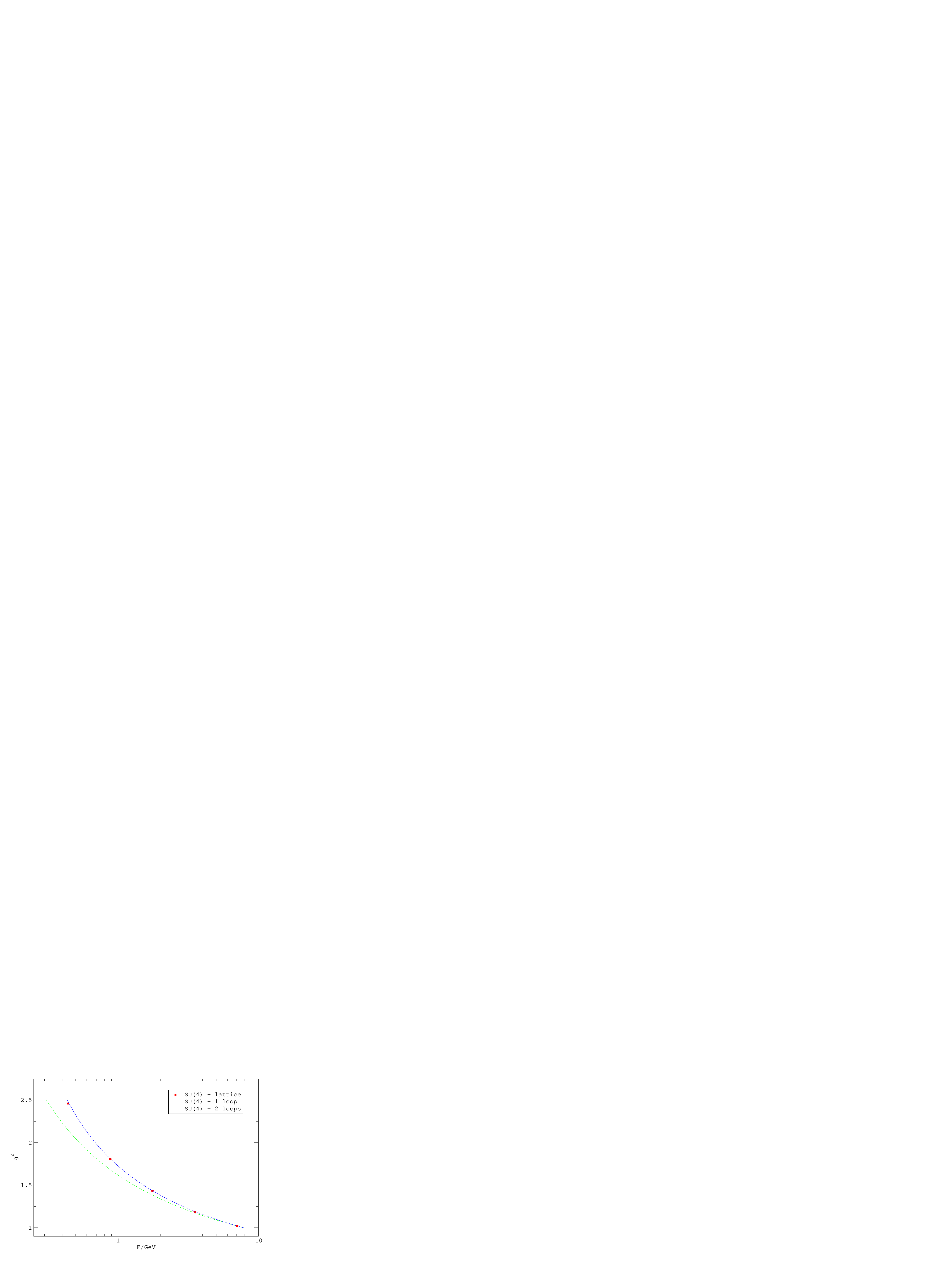}
\end{center}
\caption{Results of lattice simulations for the running coupling in the SU(4) theory, together with the one and two loop perturbative predictions.}
\label{fig:coupling}
\end{figure}

Figure \ref{fig:coupling} shows the lattice data plotted alongside the perturbative evolution predicted by the beta function. The data is thus accurately described by two-loop perturbation theory down to energy scale of the order of $\sqrt{\sigma}$.
\subsection*{Scaling with $N$}
\label{sec:scaling}
It is expected that for large-$N$ the running of the t'Hooft coupling $\bar g^2N$ is independent of $N$, so that a universal curve describes the running for any~$N$. Previous lattice simulations have hinted that this holds also at energy scales where one would expect large-$N$ perturbation theory to break down~\cite{Lucini:2001ej}. We are now in a position to test this using the results herein, together with those of SU(2) and SU(3) in \cite{Luscher:1992zx,Luscher:1993gh}.

\begin{figure}
\begin{center}
\includegraphics[scale=0.45]{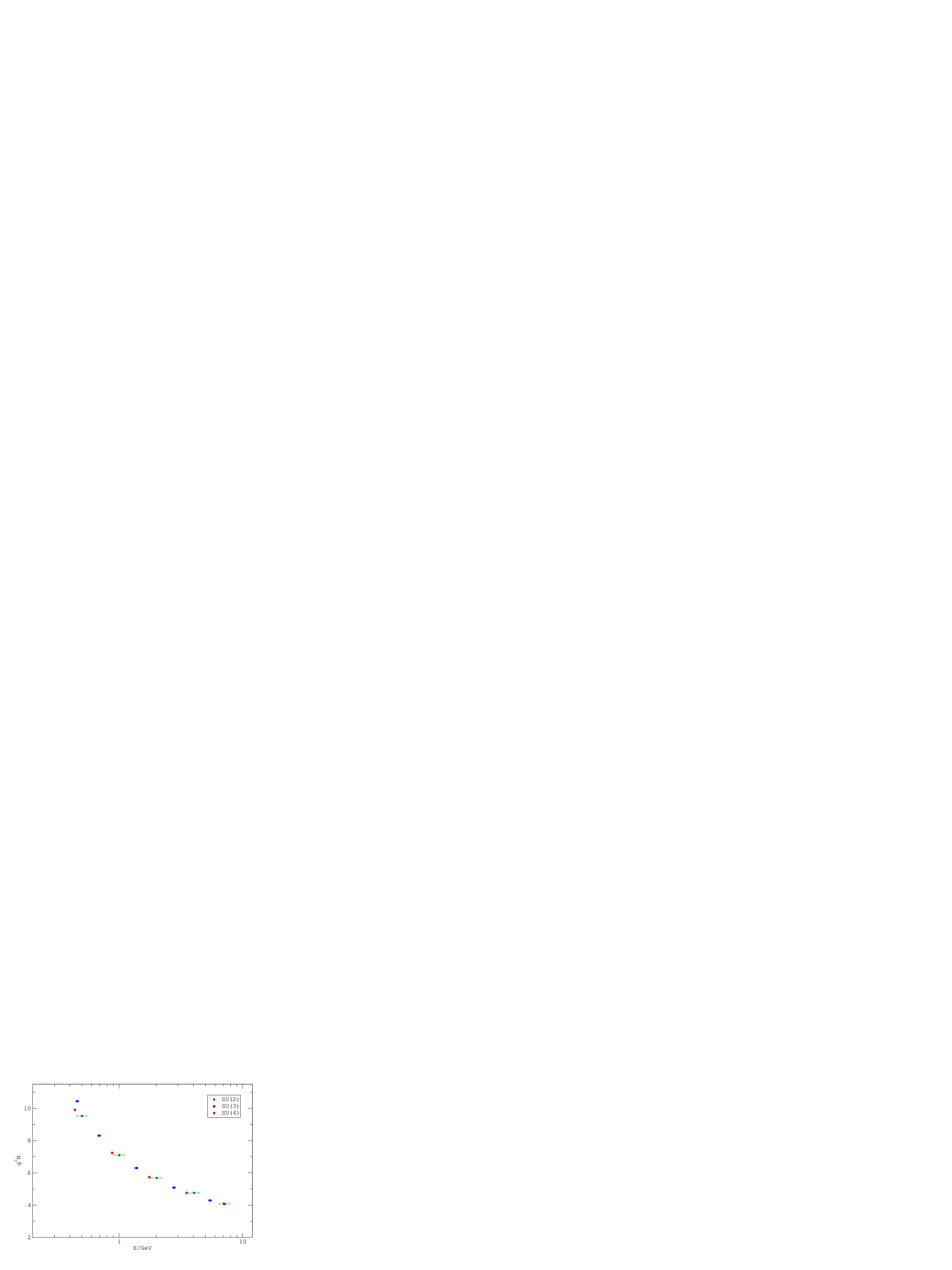}
\end{center}
\vspace{-3mm}
\caption{Running of the t'Hooft coupling $\bar g^2N$ for $N=2,3,4$.}
\label{fig:scaling}
\end{figure}

As for the case of SU(4), we fix the physical scale of the SU(2) and SU(3) data using updated values of the string tension~\cite{Lucini:2004my,Lucini:2005vg}, and superimpose the results for $N=2,3,4$ (figure \ref{fig:scaling}). The points fall on a single curve within the errors, suggesting that the prediction of universality is accurate even at $N=2$. Again, the errors shown correspond to the scale uncertainty introduced when fitting to the string tension and we expect this to be the biggest source of error, especially in the case of SU(2) where the range of $\beta$ for which we have accurate values of the string tension does not overlap with the range of $\beta$ covered in the Schr\"odinger functional simulation \cite{Luscher:1992zx}.
The corresponding $\Lambda_{SF}$ were found to be
\begin{equation}
\Lambda_{SF}(N=2)=110\pm10\operatorname{MeV},\qquad\Lambda_{SF}(N=3)=103\pm3\operatorname{MeV}.
\end{equation}
From general theoretical considerations, we expect that the leading order corrections to the universal behaviour are of order $1/N^2$. In fact, our data are consistent with this expectation. Though a fuller analysis of the errors should be carried out at this stage, we tentatively find that, in the range of $N$ considered, $\Lambda_{SF}(N)$ is well approximated by
\begin{equation}
\frac{\Lambda_{SF}(N)}{\sqrt\sigma}=0.23+\frac{0.12}{N^2}
\end{equation}
\section{Conclusions}
\label{sec:conclusions}
We have generalised the Schr\"odinger functional to SU(4), and used it to produce data for the running coupling from 7GeV down to the scale of $\sqrt\sigma$, effectively linking the perturbative to the non-perturbative regime. The result is similar to the previous SU(2) and SU(3) calculations already performed in that the data is well approximated by perturbation theory down to energies of the order of the string tension. Of course, this should not be taken to mean that perturbation theory correctly accounts for all phenomena down to this energy, as the coupling defined through the Schr\"odinger functional may simply be an exceptional case.

Similarly, the t'Hooft coupling $\bar g^2N$ is seen to be a universal function of $E$, only weakly dependent on $N$ even down to $N=2$, and we have extracted the $N$ dependence of $\Lambda_{SF}$ to leading order in $1/N^2$. The large-$N$ expectation of universality thus holds true down to energies of order $\sqrt\sigma$, though we are unable to make any certain claims as to whether universality really extends to the non-perturbative level, since perturbation theory itself proved adequate in the ranges of energy studied.

\end{document}